\documentclass[prd, twocolumn, nofootinbib, floatfix]{revtex4-1}

\usepackage{amsmath}
\usepackage{graphicx}
\usepackage{dcolumn}
\usepackage{bm}
\usepackage{epsfig}
\usepackage{amssymb,latexsym,mathrsfs}
\usepackage{graphicx}
\usepackage{color}
\usepackage{hyperref}

\hypersetup{
    colorlinks=true,
    linkcolor=red,
    citecolor=blue,
} 

\newcommand{\be}{\begin{equation}}
\newcommand{\ee}{\end{equation}}
\newcommand{\bs}{\begin{split}} 
\newcommand{\bea}{\begin{eqnarray}}
\newcommand{\eea}{\end{eqnarray}}

\newcommand{\om}{\Omega}

\newcommand{\omm}{\Omega_m}

\newcommand{\gm}{G_{\rm matter}} 
 
\newcommand{\gl}{G_{\rm light}} 
\newcommand{\geff}{G_{\rm eff}}

\newcommand{\al}{\alpha} 
 
\newcommand{\dl}{\delta} 

\newcommand{\fsig}{f\sigma_8}

\begin{document}

\title{No Slip Gravity} 

\author{Eric V.\ Linder${}^{1,2}$} 
\affiliation{${}^1$Berkeley Center for Cosmological Physics \& Berkeley Lab, 
University of California, Berkeley, CA 94720, USA\\ 
${}^2$Energetic Cosmos Laboratory, Nazarbayev University, 
Astana, Kazakhstan 010000}

\begin{abstract} 
A subclass of the Horndeski modified gravity theory we call No Slip 
Gravity has particularly interesting properties: 1) a speed of gravitational 
wave propagation equal to the speed of light, 2) equality between the 
effective gravitational coupling strengths to matter and light, 
$\gm$ and $\gl$, hence no slip between the metric potentials, yet 
difference from Newton's constant, and 3) suppressed growth 
to give better agreement with galaxy clustering observations. We explore 
the characteristics and implications of this theory, and project observational 
constraints. 
We also give a simple expression for the ratio of the 
gravitational wave standard siren distance to the photon standard 
candle distance, in this theory and others, and enable a direct comparison 
of modified gravity in structure growth and in gravitational waves, an 
important crosscheck. 
\end{abstract}

\date{\today} 

\maketitle

\section{Introduction} 

Modified gravity theories are becoming increasingly tested by current 
and forthcoming observations. A main motivation for considering alternatives 
to general relativity is the observation of cosmic acceleration, but 
modified gravity such as scalar-tensor theories have other degrees of 
freedom that allow scalar (density) and tensor (gravitational wave) 
perturbations to behave in ways not purely governed by the expansion, 
unlike general relativity. 

Recent measurement of the speed of propagation of gravitational waves 
from GW170817 relative to its electromagnetic counterpart GRB170817A 
\cite{ligo} severely limit its deviation from the speed of light, 
strongly disfavoring theories that generically predict a difference 
\cite{1509.08458,prl1,prl2,prl3,prl4,1711.04825}. This leaves only a 
subset out of the Horndeski class of gravity, one of the most general 
scalar tensor theories with second order field equations, a key testing 
ground for cosmology. 

Interestingly, in the remaining theories there is a unique subclass 
that makes another definite observational prediction: that another 
signature of deviation from general relativity, the slip between 
metric potentials, should vanish and yet the gravity theory does not 
reduce to general relativity. 
Indeed, discussion of the slip itself \cite{1210.0439,1305.0008} and 
the relation between slip and the speed of 
propagation of gravitational waves has highlighted important connections 
between the scalar and tensor sectors of modified gravity 
(e.g.\ see \cite{1406.7139,1407.8184,1612.02002}). 
We explore the characteristics and 
implications of this subclass, No Slip Gravity. 

In Sec.~\ref{sec:theory} we define this theory in terms of the 
property functions, the equivalent effective field theory functions, 
and the Horndeski Lagrangian functions. We demonstrate the conditions 
for stability of the theory in Sec.~\ref{sec:stable}, and give 
specific model examples. Section~\ref{sec:obs} ties this to observations 
of growth of structure, for current data and projecting constraints from 
future cosmic structure surveys. We conclude in Sec.~\ref{sec:concl}.

\section{No Slip Gravity Theory} \label{sec:theory} 

The effects of gravity on observations of cosmic matter and light can 
fruitfully be described (in the subhorizon, quasistatic limit; see 
Sec.~\ref{sec:stable}) by modified Poisson equations relating the 
time-time metric potential $\psi$ and space-space metric potential 
$\phi$ (in Newtonian gauge) to the matter perturbations. These equations 
are 
\bea 
\nabla^2\psi&=&4\pi G_N\delta\rho\times \gm\\ 
\nabla^2(\psi+\phi)&=&8\pi G_N\delta\rho\times\gl \ , 
\eea 
where the first equation governs the growth of structure, with a 
gravitational strength $\gm$, and the second governs the deflection 
of light, with a gravitational strength $\gl$. 

The offset between $\gm$ and $\gl$, or $\psi$ and $\phi$, is referred 
to as the gravitational slip, with 
\bea 
\bar\eta\equiv\frac{\gm}{\gl}&=&\frac{2\psi}{\psi+\phi}=\frac{2\eta}{1+\eta}\\ 
\eta\equiv\frac{\psi}{\phi}&=&\frac{\gm}{2\gl-\gm}=\frac{\bar\eta}{2-\bar\eta} 
\ . 
\eea 
Note that when $\bar\eta=1$ then $\eta=1$ and the converse. This corresponds 
to vanishing slip. 

The expressions for $\gm$, $\gl$, and slip in Horndeski gravity, or the 
equivalent effective field theory (EFT) approach, are given in, e.g.,  
\cite{bell,gubitosi,lsw}. Imposing that the speed of propagation of 
gravitational 
waves equals the speed of light, $c_T=1$ or $\al_T\equiv c_T^2-1=0$, 
simplifies the equations and we find a simple criterion for no slip: 
\bea 
\al_B&=&-2\al_M\\ 
m_0^2\dot\Omega&=&\bar M_1^3\\ 
G_{4\phi}&=&-XG_{3X} \ . 
\eea 
Here the second two equations give the equivalent conditions in the EFT 
and Horndeski function approaches; we work in the $\alpha_i$ parameter 
approach but show the others for convenience. The first equation relates 
the property functions $\al_i(t)$, with $\al_B$ the braiding function 
mixing scalar and tensor properties and $\al_M$ the running of the Planck 
mass. The second relates two EFT functions, the conformal factor 
$m_0^2\Omega(t)$ multiplying the Ricci scalar and the $\bar M_1^3(t)$ 
function multiplying the product of the trace of the extrinsic curvature and 
the lapse function, $\dl K\,\dl g^{00}$. The third relates different 
terms in the Horndeski Lagrangian, where a subscript $\phi$ denotes a 
derivative with respect to the scalar field $\phi$ (not the metric potential) 
and a subscript $X$ denotes a derivative with respect to the scalar kinetic 
energy $X=-g^{\mu\nu}\partial_\mu\phi\partial_\nu\phi/2$. 

The implications of No Slip Gravity for the gravitational strengths are 
quite simple also: 
\be 
\gm=\gl=\frac{m_p^2}{M_\star^2} \ , 
\ee 
where $m_p$ is the Planck mass in general relativity, and 
$M_\star^2\,(=m_0^2\Omega)$ is the effective, time dependent Planck mass 
in the modified gravity. Note that $\al_M=(1/H)d\ln M_\star^2/dt$. 
Thus matter growth 
and light propagation are modified in step, as time dependent effects, 
and are distinct from general relativity. Such a simple theory provides 
an excellent test ground for the ability of future cosmic surveys to 
look for deviations from general relativity and constrain their magnitude.

\section{Viability Conditions} \label{sec:stable}

For any theory of gravity it is important to make sure it has a firm 
foundation, without pathologies or instabilities. In terms of the property 
functions, the condition for no ghosts is \cite{bell} 
\be 
\al_K+\frac{3}{2}\al_B^2>0 \ . 
\ee 
Since gravitational waves propagate at the speed of light the tensor 
sector is also free of ghosts. 
The quasistatic approximation is discussed in detail in 
\cite{1108.4242,bell,1503.06831}; this basically reduces to 
$\alpha_B\,k/(aH)\gg1$, satisfied for modes well within the horizon if 
$\alpha_B$ is not so small that we would see no modified gravity effect 
anyway. 

One must also check for instabilities in the scalar sector, with the 
stability condition $c_s^2\ge0$. Expressions for the 
sound speed $c_s$ are given in the property function and EFT 
approaches in, e.g., \cite{bell,1411.3712,1605.06102,1601.04064}. 
Within the property function approach, imposing $\al_T=0$ yields 
\be 
c_s^2=\frac{(1-\al_B/2)[H^2(2\al_M+\al_B)-2\dot H]+H\dot\al_B-\rho_m-p_m}{H^2(\al_K+3\al_B^2/2)} 
\ee 
where $H$ is the Hubble parameter, and $\rho_m$ and $p_m$ are 
the matter energy density and pressure. 
We are free to choose the background expansion separately from the 
perturbation terms, and we adopt a $\Lambda$CDM background, in good 
agreement with current observations. This cancels terms involving the 
matter fields and $\dot H$. 

Within the EFT approach, 
\bea  
c_s^2 \propto \frac{-3\dot\Omega}{H\Omega}&+&\frac{\bar M_1^3}{H m_0^2\Omega}\\  
&+&\frac{2\dot\Omega/(H\Omega)+\ddot\om/(H^2\om)+\dot{\bar M}_1^3/(H^2 m_0^2\om)}{1+\dot\om/(2H\om)+\bar M_1^3/(2H m_0^2\om)} \ ,\notag  
\eea 
where we omit the denominator, which must be positive by the no ghost 
condition. 

For No Slip Gravity with $\al_B=-2\al_M$ or $\bar M_1^3=m_0^2\dot\om$ 
the stability condition takes the simple, equivalent forms 
\bea 
(H\al_M)\,\dot{}&<&0 \label{eq:stable}\\ 
(\dot\om/\om)\dot{}&<&0 \ , 
\eea 
in the property function and EFT approach respectively. 
Using the definitions of $\al_M$ and $m_0^2\om$ in terms of $M_\star^2$ 
shows these conditions are indeed equivalent. Indeed, they are just 
\be 
\frac{d^2\ln M_\star^2}{dt^2}<0 \ . 
\ee 

We now have a one function theory, similar to $f(R)$ gravity (which has 
$\al_B=-\al_M$ and so does have slip), for a given background expansion. 
Note that the stability condition is quite restrictive. If we write 
\be 
(H\al_M)\dot{}=\dot H \al_M+H\dot\al_M<0 \ , 
\ee 
then we can see that since $\dot H<0$ for all times, and $\al_M\to0$ as 
the universe approaches its de Sitter asymptote, that stability must 
break down at some time when $\al_M<0$. That is, negative $\al_M$ must 
climb back to zero, either to cross to positive $\al_M$ or to reach 
the de Sitter limit, giving $\dot\al_M>0$ and this will violate stability. 
In the early universe we want gravity to restore to general relativity, 
with $\al_M=0$. So the simplest viable form is a ``hill'' in $\al_M$, 
where it is always positive or zero. 
Note that a form such as 
$\al_M=\mu\Omega_\Lambda$, proportional to the effective dark 
energy density $\Omega_\Lambda(t)$, is immediately ruled out as it 
does not approach zero at late times as required, while 
$\al_M=\mu\Omega_\Lambda(1-\Omega_\Lambda)$ is unstable. 

For the one function determining the theory, we can choose either 
$M_\star^2$ or $\al_M$. Note that in No Slip Gravity the condition 
$\al_B=-2\al_M$ determines that 
\be 
\gm=\gl=\frac{m_p^2}{M_\star^2} \ , \label{eq:gmm} 
\ee 
an extraordinarily simple result. In the early universe, we take 
$M_\star^2\to m_p^2$ so the gravitational strength agrees with general 
relativity. In the asymptotic future, $M_\star^2$ freezes to 
its de Sitter value, and so does the gravitational strength. 

Let us explore parametrizing $M_\star^2(t)/m_p^2$. We need it to transition 
from unity in the past to some constant value $1+\mu$ in the future. Other 
than that, a wide variety of functional forms is possible. We take the 
following as purely illustrative examples satisfying these conditions. 
One such simple function is the e-fold or $1+\tanh$ form 
\bea 
\frac{M_\star^2}{m_p^2}&=&1+\frac{\mu}{1+e^{-\tau(\ln a-\ln a_t)}}=
1+\frac{\mu}{1+(a/a_t)^{-\tau}}\\ 
&=&1+\mu\,\frac{1+\tanh[(\tau/2)\ln(a/a_t)]}{2} 
\ , \label{eq:mefold} 
\eea  
where $a$ is the cosmic scale factor. 
Here $\mu$ gives the amplitude of the transition, $a_t$ the scale factor 
when it occurs, and $\tau$ its rapidity. For this form, the stability 
condition requires $0<\tau\le 3/2$. 

\begin{figure}[htbp!]
\includegraphics[width=\columnwidth]{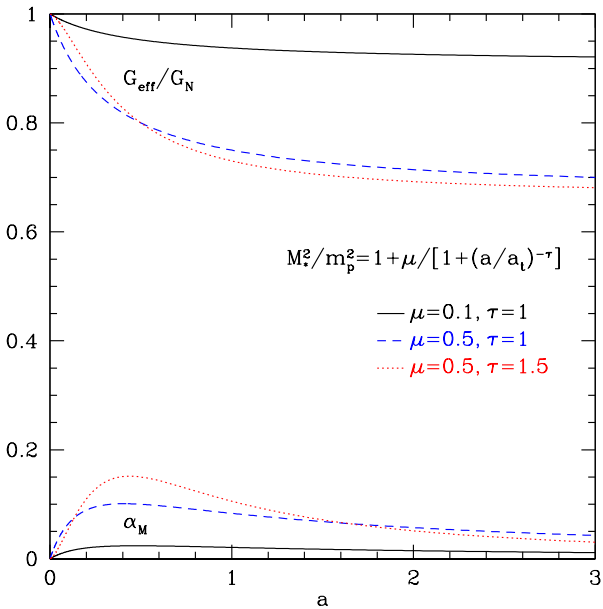} 
\caption{
A model for $M_\star^2(t)$ determines the theory. Here we plot 
$\geff(a)/G_N=\gm=\gl$ and $\al_M$ for various values of the model amplitude $\mu$ 
and evolution rapidity $\tau$, for fixed transition time $a_t=0.5$. 
} 
\label{fig:mefold} 
\end{figure}

Using $\al_M=d\ln M_\star^2/d\ln a$ we find 
\be 
\al_M=\left[1+\frac{\mu}{1+e^{-\tau(\ln a-\ln a_t)}}\right]^{-1} 
\frac{\tau\mu e^{-\tau(\ln a-\ln a_t)}}{\left[1+e^{-\tau(\ln a-\ln a_t)}\right]^2} \ . \label{eq:alefold} 
\ee 
This vanishes at early and late times as required, and reaches a 
maximum in the vicinity of $a_t$, with amplitude $\al_M\approx \mu\tau/4$. 
The results for the evolutions of $\gm$ and $\al_M$ are presented in 
Fig.~\ref{fig:mefold}.

An alternate approach is to parametrize $\al_M$, and derive $M_\star^2$ 
by integration. Recalling that we want $\al_M$ to be a hill, vanishing 
at early and late times, we adopt the (again, purely illustrative) form 
\be 
\al_M=A\left(1-\tanh^2[(\tau/2)\ln(a/a_t)]\right)= 
\frac{4A\,(a/a_t)^\tau}{[(a/a_t)^\tau+1]^2} \ . 
\ee 
Again, for this form the stability condition requires $0<\tau\le 3/2$. 
This is easy to understand since at early times $\al_M\sim a^{\tau}$, 
as for the form of Eq.~(\ref{eq:alefold}), and during matter domination 
the Hubble parameter $H\sim a^{-3/2}$, so 
$(H\al_M)\,\dot{}\sim (a^{\tau-3/2})\,\dot{}\le0$ requires $\tau\le3/2$. 

\begin{figure}[htbp!]
\includegraphics[width=\columnwidth]{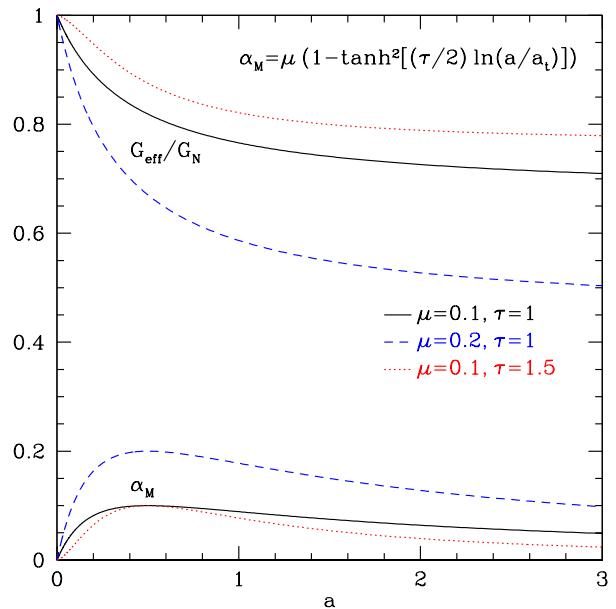} 
\caption{
A model for $\al_M(t)$ determines the theory. Here we plot 
$\geff(a)/G_N=\gm=\gl$ and $\al_M$ for various values of the model amplitude $A$ 
and evolution rapidity $\tau$, for fixed transition time $a_t=0.5$. 
} 
\label{fig:a2} 
\end{figure}

For this approach, $\al_M$ reaches a maximum of $A$ at $a_t$, and 
vanishes in the past and future. We can write the corresponding 
$M_\star^2$ analytically as 
\be 
\frac{M_\star^2}{m_p^2}=e^{(2A/\tau)(1+\tanh[(\tau/2)\ln(a/a_t)])} \ . 
\ee 
Note that the form in Eq.~(\ref{eq:mefold}) is basically the first order 
expansion of this. In the past, this goes to unity, and in the future 
it goes to a constant $e^{4A/\tau}$. 
The results for the evolutions of $\gm$ and $\al_M$ are presented in 
Fig.~\ref{fig:a2}.

A major consequence of the stability condition that we have seen in 
Figs.~\ref{fig:mefold} and \ref{fig:a2} is that the gravitational strength 
is weaker than in general relativity! This follows, independent of the 
functional forms, because stability 
requires that $\al_M\ge0$. Recall that 
\be 
\ln\frac{M_\star^2(a)}{m_p^2}=\int_0^a d\ln\tilde a\,\al_M(\tilde a) \ . 
\ee 
Since $\al_M(a)\ge0$ for all $a$, then $M_\star^2(a)/m_p^2\ge1$, and 
so by Eq.~(\ref{eq:gmm}), $\gm=\gl\le1$. 

This is quite unusual for scalar-tensor theories, which generically 
increase the strength of gravity. However, it arises in No Slip Gravity 
due to the strength of the braiding, which mixes the scalar sector into 
the tensor sector (cf.\ \cite{1008.0048,1609.01272,1712.00444}). 

As we see in the next section, this weakening of gravity, despite the 
presence of cosmic acceleration, has important and potentially beneficial 
consequences for observations.

\section{Observational Constraints} \label{sec:obs}

Recently, \cite{mglowz} showed that binned values of $\gm$ gave highly 
accurate reconstructions of the observable growth quantities. 
Since in the No Slip Gravity case $\gm=(m_p^2/M_\star^2)$ then one can 
directly read off from the binned $\gm$ the central quantity of the 
theory, 
\be 
\al_M=-\Delta\ln\gm/\Delta\ln a \ . 
\ee 

We have verified that for both the $M_\star^2(a)$ and $\al_M(a)$ models 
the binned approach delivers the redshift space distortion growth rate 
observable $\fsig$ to 0.2\% accuracy. 

A closely related observational relation concerns the gravitational 
wave (GW) standard siren distance. Following  
\cite{1710.04825,1711.03776} we see that the GW strain amplitude 
\bea 
h &=& h^{GR} e^{-(1/2)\int_{\rm em}^{\rm obs} d\ln a\,\alpha_M(a)}  
= h^{GR} e^{-(1/2)\int_{\rm em}^{\rm obs} d\ln M^2_\star(a)} \\ 
&=& h^{GR} \left[\frac{M^2_{\star,{\rm em}}}{M^2_{\star,{\rm obs}}}\right]^{1/2} \ . 
\eea 
Since the strain is inversely proportional to the standard siren 
luminosity distance, one has\footnote{During 
the late stages of this work, \cite{1712.08623} 
appeared with an equivalent expression.} 
\be 
d_{L,GW}(a)=d_L^{GR}(a)\,
\left[\frac{M^2_{\star}(a=1)}{M^2_{\star}(a)}\right]^{1/2}\ . 
\ee 

This is a quite general expression for Horndeski gravity and some 
other theories. 
Note in particular that the photon luminosity distance is simply $d_L^{GR}$ 
so a comparison of the GW standard siren distance and the photon standard 
candle distance gives a simple test of gravity. 
Thus one can in principle measure the evolution of 
$M_\star(a)$; the running $\al_M$ would require a derivative of 
noisy data. For No Slip Gravity we have the further simplification 
that 
\be 
d_{L,GW}(a)=d_L^{GR}(a)\,\left[\frac{\gm(a)}{\gm(a=1)}\right]^{1/2}\ , 
\ee 
and one could compare the modified gravity derived from GW in the tensor 
sector to that from growth of structure in the scalar sector. 
(After this article first appeared, \cite{1712.08108} showed the same 
relation holds in a theory of nonlocal gravity.) 

\begin{figure}[htbp!]
\includegraphics[width=\columnwidth]{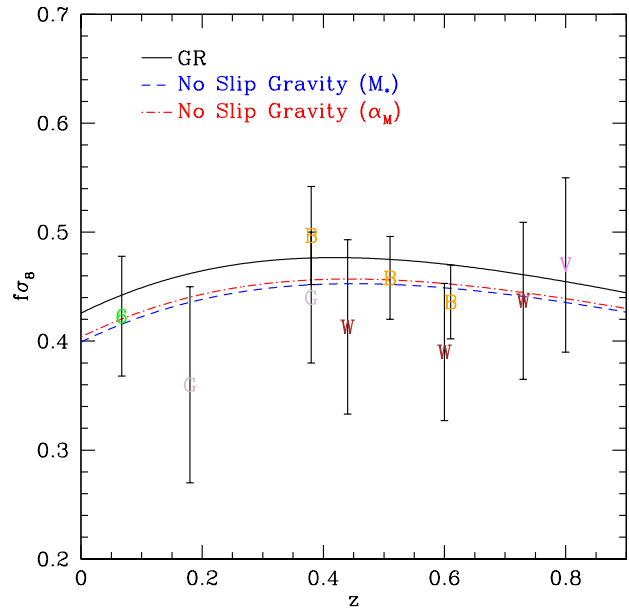} 
\caption{
Current measurements of the cosmic structure growth rate $\fsig$ are 
compared with the general relativity prediction for the Planck cosmology 
($\omm=0.31$; solid black curve) and the No Slip Gravity models of 
$M_\star$ (dashed blue) and $\al_M$ (dot dashed red) functions. The data 
points come from 6dFGRS (6; \cite{6dF}), GAMA (G; \cite{gama}), BOSS 
(B; \cite{alam}), WiggleZ (W; \cite{wigglez}), and VIPERS (V; \cite{vipers}). 
} 
\label{fig:fscurve} 
\end{figure}

Returning to growth observables, galaxy redshift surveys already have a 
slew of measurements of the growth rate quantity $\fsig$. We can apply 
our illustrative forms to examine the impact on this observable; we 
emphasize this is meant as a demonstration of principle regarding 
suppression of growth and not a fully quantitative likelihood analysis. 
Figure~\ref{fig:fscurve} compares the predictions of No Slip Gravity, 
where we use the exact solution of growth, with 
the cosmic expansion fixed to the best fit Planck cosmology (i.e.\ flat 
$\Lambda$CDM with $\omm=0.31$), to a compendium of current observations.

The fits of the two representative models of No Slip Gravity, employing a 
motivated functional form for $M_\star^2(a)$ and $\al_M(a)$ respectively, 
appear to improve over the concordance cosmology. 
Recall they have the same expansion history as the Planck 
cosmology, and so will fit distance data as well as the concordance, general 
relativity cosmology. Figure~\ref{fig:fscurve} illustrates they 
provide fits more in accord with the growth rate data coming 
from redshift space distortion measurements, however. 
We find that current observations are in agreement with the $M_\star^2$ model 
with $\mu=0.1$ or the $\al_M$ model with $A=0.03$, both with transition 
time $a_t=0.5$ and $\tau=1.5$. Again, these numbers are meant to give an 
indication of the characteristics, not a detailed analysis. 

\begin{figure}[htbp!]
\includegraphics[width=\columnwidth]{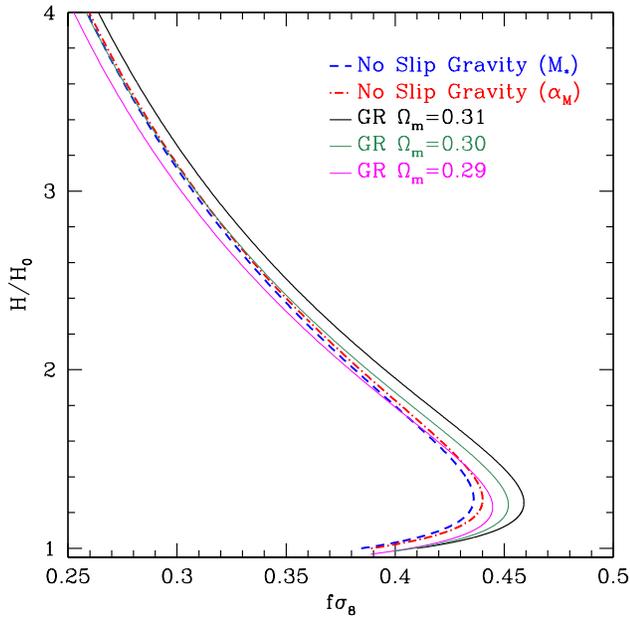} 
\caption{
Using the conjoined growth-expansion approach illustrates the distinction 
between modified gravity and general relativity, in terms of the evolution 
in the $\fsig(z)$ vs $H(z)/H_0$ plane. The behavior of the No Slip Gravity 
models of $M_\star$ (dashed blue) and $\al_M$ (dot dashed red) functions 
have characteristic deviations from the general relativity predictions for 
the Planck cosmology ($\omm=0.31$; solid black curve) and other background 
cosmologies ($\omm=0.3$ thin solid green and $\omm=0.29$ thin solid magenta). 
Note that $H_0$ here is that of the $\omm=0.31$ case and other cases are 
scaled to preserve the CMB sound horizon. 
} 
\label{fig:conjoin} 
\end{figure}

We can further highlight the deviation from general relativity by 
employing the conjoined expansion and growth history visualization of 
\cite{conjoin}. Figure~\ref{fig:conjoin} illustrates that the modification 
of gravity is distinct from a change in the background cosmological model. 
Recall that for the No Slip Gravity models we adopted the Planck cosmology 
of flat $\Lambda$CDM with $\omm=0.31$, but we see the modified gravity 
conjoined growth-expansion history in terms of $\fsig$ vs $H$ does not 
lie along the general relativity curves. While one can change the background 
to match the modified gravity prediction over a narrow range of redshifts, 
the modified gravity model has its own characteristic behavior.

Next we consider the leverage of next generation observations, such as 
from the Dark Energy Spectroscopic Instrument (DESI \cite{desi}), with 
percent level measurements of $\fsig$ to test gravitation theory. We 
carry out a Fisher information analysis following the approach of 
\cite{1703.00917} in testing early modified gravity. The data is taken 
to be future measurements of $\fsig$ in 18 redshift bins over $z=0.05$--1.85 
as projected by \cite{desi}. Only linear modes are used, out to 
$k_{\rm max}=0.1\,h$/Mpc. We include a Gaussian prior on the matter density 
$\omm$ of $0.01$ to represent external data such as Planck CMB measurements. 

For the gravity model we take the fit parameters as exhibited in 
Fig.~\ref{fig:fscurve}, for the two cases. In each case we fix $a_t=0.5$ 
as a reasonable transition time and $\tau=1.5$ as the maximum allowed 
rapidity. Constraints weaken for early or late transitions, and slow 
ones, due to parameter degeneracies so we present an optimistic scenario 
for searching for modifications to gravity; we seek an indication of 
the sensitivity, not meaning this as a detailed likelihood fit. 
We fit for the matter density 
and amplitude of the deviation from general relativity, either $\mu$ in 
the $M_\star^2$ model or $A$ in the $\al_M$ model. Both correspond to the 
maximum deviation over time of the functions from the general relativity 
limit. 

\begin{figure}[htbp!]
\includegraphics[width=\columnwidth]{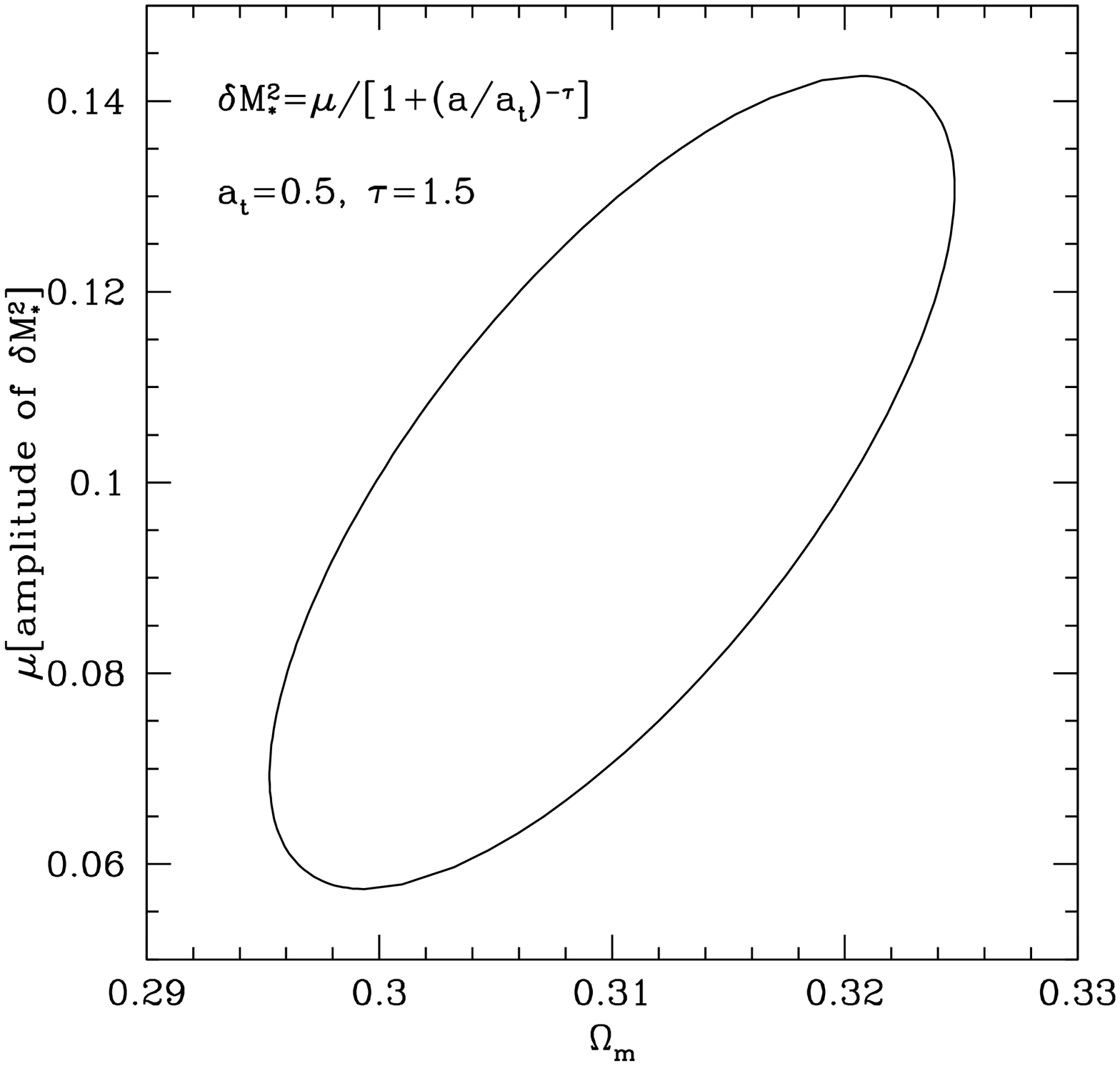} 
\caption{
68\% joint confidence level constraints from future DESI redshift space 
distortion data on the No Slip Gravity model with 
$M_\star^2/m_p^2=1+\mu/[1+(a/a_t)^{-\tau}]$ are shown in the $\omm$--$\mu$ 
plane, centered on the current best fit values. This can give a $\sim3\sigma$ 
detection of $\mu$, i.e.\ a deviation from general relativity. 
} 
\label{fig:ellommu} 
\end{figure}

\begin{figure}[htbp!]
\includegraphics[width=\columnwidth]{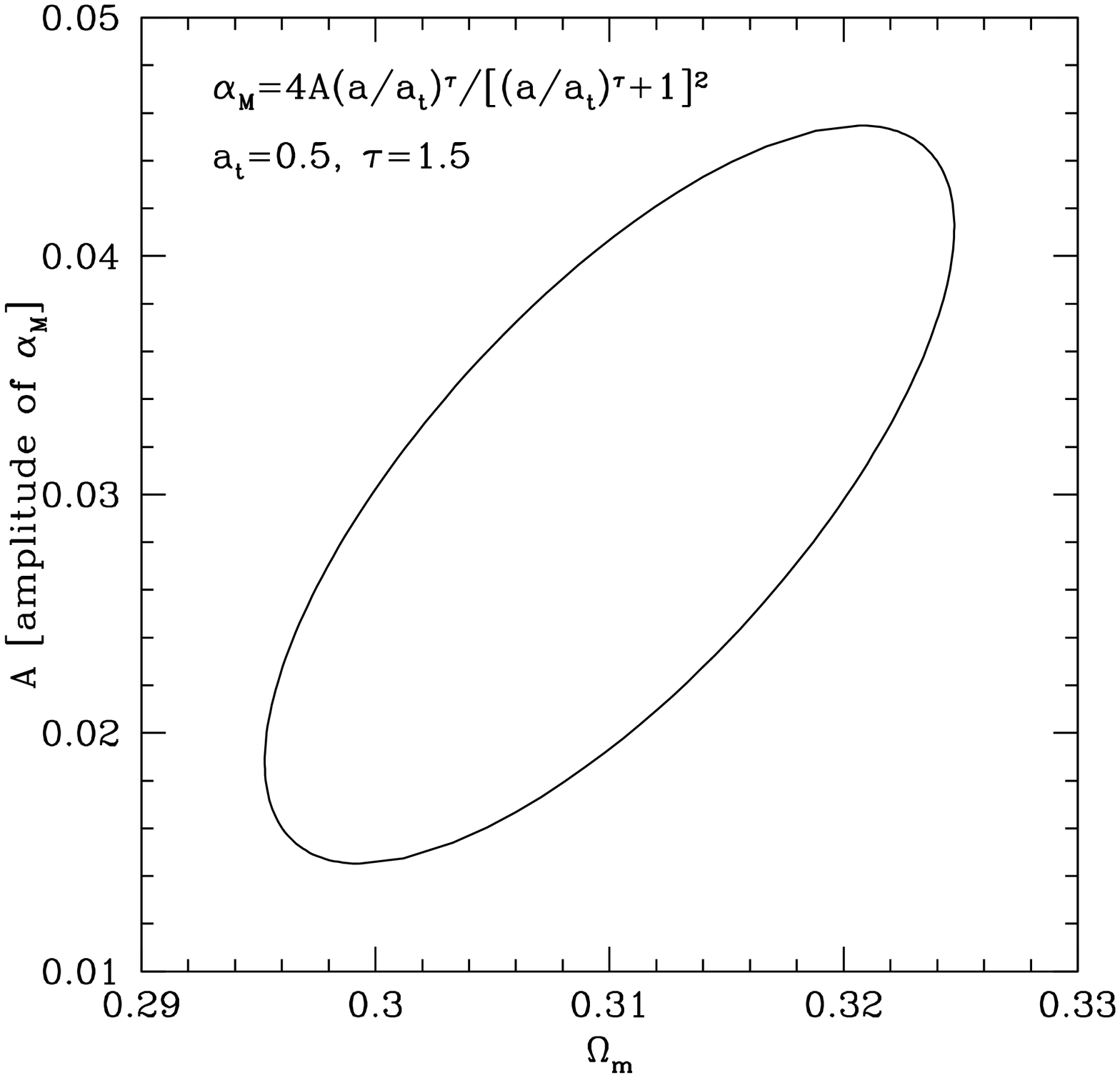} 
\caption{
68\% joint confidence level constraints from future DESI redshift space 
distortion data on the No Slip Gravity model with 
$\al_M=4A(a/a_t)^\tau/[(a/a_t)^\tau+1]^2$ 
are shown in the $\omm$--$A$ 
plane, centered on the current best fit values. This can give a $\sim3\sigma$ 
detection of $A$, i.e.\ a deviation from general relativity. 
} 
\label{fig:elloma} 
\end{figure}

Figures~\ref{fig:ellommu} and \ref{fig:elloma} show the results. The 
marginalized constraints on the modified gravity amplitudes are 
$\mu=0.1\pm0.028$ and $A=0.03\pm 0.010$ respectively. Next 
generation data could see signatures of modification of gravity 
at the $\sim3\sigma$ level in either model, if either model is correct 
and redshift space distortion data continue to lie along the current best 
fit.

\section{Conclusions} \label{sec:concl} 

The constraint on the speed of propagation of gravitational waves from 
GW170817/GRB170817A severely limited theories of modified gravity. Many 
conformal scalar-tensor theories still remain but only two carry 
particularly significant observational implications. The class of $f(R)$ 
gravity predicts that $\gl=1$ (assuming $f_R\ll1$ as required from 
solar system and astrophysical tests). Here we presented the other -- 
No Slip Gravity -- which makes the gravitational slip vanish so 
$\gm=\gl$, though they can still differ from general relativity. 

(Since $f(R)$ has $\al_B=-\al_M$ and $\gl=1$, and No 
Slip Gravity has $\al_B=-2\al_M$ and $\eta=1$, one might imagine studying 
an interpolation (or extrapolation) $\al_B=-R\al_M$ but there is no 
equivalent physics motivation.) 

No Slip Gravity is a simple one function theory, and the form of the 
function with time is strongly constrained by the stability condition 
$c_s^2\ge0$. In particular this implies that the running of the Planck 
mass $\al_M\ge0$ at all times. We presented two representative models, 
one in terms of a viable Planck mass function $M_\star^2(a)$ and one 
in terms of a viable running $\al_M(a)$. 

Unlike many scalar-tensor theories, No Slip Gravity makes the definite 
observational prediction that gravity should be weaker than in general 
relativity, despite giving cosmic acceleration. We showed that this is 
in excellent agreement with current redshift space distortion data 
measuring the cosmic structure growth rate $\fsig(a)$, better than 
general relativity for the Planck cosmology. Potentially it could also 
inform the tension on the weak lensing quantity 
$S_8=\sigma_8(\omm/0.3)^{0.5}$ 
\cite{1606.05338,1707.06627,1708.01530,1708.01538} and the 
value of $E_G$ lower than general relativity \cite{1711.10999} 
though we leave that for future study. 

We also gave a simple expression for the ratio of the 
gravitational wave standard siren distance to the photon standard 
candle distance, in this theory and others. This enables a comparison 
of modified gravity in structure growth and in gravitational waves, 
an important crosscheck. 

Next generation galaxy redshift survey data could distinguish between 
general relativity and No Slip Gravity at the $\sim3\sigma$ level, if 
the percent level measurements of $\fsig$ lie along the current best 
fit. Next generation imaging surveys, such as Euclid and LSST, could 
test the prediction of No Slip Gravity that there is no slip. Such 
tests would be an exciting development, searching for signatures of 
modified gravity that makes definite predictions.

\acknowledgments 

This work is supported in part by the Energetic Cosmos Laboratory and by 
the U.S.\ Department of Energy, Office of Science, Office of High Energy 
Physics, under Award DE-SC-0007867 and contract no.\ DE-AC02-05CH11231.

\appendix 

\section{No Running} 

One might notice that another way to obtain zero slip (assuming $\al_T=0$) 
is to impose $\al_M=0$. This gives 
\bea 
\gm&=&\gl=\frac{2\al_B+2\al'_B}{(2-\al_B)\al_B+2\al'_B} \\ 
c_s^2&\propto&(H\al_B)\dot{}+[1-(\al_B/2)]\,\al_B H^2 > 0 \ . 
\eea 
This restores to general relativity in the early universe when $\al_B\ll1$. 
(Also see \cite{1711.04825}.) 
In the late time de Sitter limit, the gravitational strength reaches 
$G_{\rm eff,dS}=1/(1-\al_B/2)$. Stability requires that $\al_B>0$ in 
this limit. However, at early times one can obtain stability with either 
sign of $\al_B$. For example, $\al_B$ can deviate from 0 to positive 
values, and continue to its de Sitter asymptote. This will be stable at 
early times if $\al'_B/\al_B>1/2$. If $\al_B$ initially deviates to negative 
values, it can be stable at early times with $\al'_B/\al_B<1/2$, but 
will force $\geff$ negative at some time later before $\al_B$ 
crosses zero on the way to its positive de Sitter asymptote. 
Thus we require $\al_B\ge0$.


\end{document}